\documentclass[
    pre,
    twocolumn,
    reprint,
    amssymb,
    amsmath,
    amsfonts,
    a4paper,
    aps,
    superscriptaddress,
    showpacs]{revtex4-1}
\usepackage{xspace, units}
\usepackage{graphicx}

\usepackage[absolute]{textpos}
\usepackage[T1]{fontenc}
\usepackage{xspace, units}
\usepackage{amssymb, amsmath, amsfonts}

\newcommand{\numdp}{\ensuremath{N}\xspace}

\newcommand{\condprob}[2]{\ensuremath{p\bigl(#1\! \mid\! #2 \bigr)}\xspace}
\newcommand{\joinprob}[2]{\ensuremath{p\bigl(#1, #2 \bigr)}\xspace}
\newcommand{\sysX}{\ensuremath{X}\xspace}
\newcommand{\sysY}{\ensuremath{Y}\xspace}

\newcommand{\sysXY}{\ensuremath{{\scriptscriptstyle \sysX \to \sysY}}\xspace}
\newcommand{\sysYX}{\ensuremath{{\scriptscriptstyle \sysY \to \sysX}}\xspace}

\newcommand{\obsX}{\ensuremath{x}\xspace}
\newcommand{\obsY}{\ensuremath{y}\xspace}
\newcommand{\stateX}[1]{\ensuremath{\obsX_{#1}}\xspace}
\newcommand{\stateY}[1]{\ensuremath{\obsY_{#1}}\xspace}

\newcommand{\symbDelay}{\ensuremath{l}\xspace}
\newcommand{\symbDim}{\ensuremath{m}\xspace}
\newcommand{\symbRank}[1]{\ensuremath{k_{i#1}}\xspace}

\newcommand{\delay}{\ensuremath{\tau}\xspace}
\newcommand{\delayS}{\ensuremath{\delay_1}\xspace}
\newcommand{\delayF}{\ensuremath{\delay_2}\xspace}
\newcommand{\delaySF}{\ensuremath{\delayS,~\delayF}\xspace}

\newcommand{\delaycplX}{\ensuremath{\Delta_\sysX}\xspace}
\newcommand{\delaycplY}{\ensuremath{\Delta_\sysY}\xspace}
\newcommand{\couplX}{\ensuremath{c_\sysXY}\xspace}
\newcommand{\couplY}{\ensuremath{c_\sysYX}\xspace}

\newcommand{\symbX}{\ensuremath{\hat \obsX}\xspace}
\newcommand{\symbY}{\ensuremath{\hat \obsY}\xspace}

\newcommand{\STE}{\ensuremath{\operatornamewithlimits{\hat T}}\xspace}
\newcommand{\DSTE}{\ensuremath{\operatornamewithlimits{\mathcal{T}}}\xspace}
\newcommand{\DSTEd}{\ensuremath{\DSTE(\delayS, \delayF)}\xspace}
\newcommand{\DSTEs}{\ensuremath{\operatornamewithlimits{\mathfrak{T}}}\xspace}

\newcommand{\SNR}{\ensuremath{\operatornamewithlimits{SNR}}\xspace}

\newcommand{\refeq}[1]{Eq.~\ref{#1}}
\newcommand{\reffig}[1]{Fig.~\ref{#1}}

\begin{document}
\title{Identifying delayed directional couplings with symbolic transfer entropy}
\author{Henning \surname{Dickten}}
  \email{hdickten@uni-bonn.de}
\author{Klaus \surname{Lehnertz}}
  \email{klaus.lehnertz@ukb.uni-bonn.de}
\affiliation{Department of Epileptology, University of Bonn, Sigmund-Freud-Stra{\ss}e~25, 53105~Bonn, Germany}
\affiliation{Helmholtz Institute for Radiation and Nuclear Physics, University of Bonn, Nussallee~14--16, 53115~Bonn, Germany}
\affiliation {Interdisciplinary Center for Complex Systems, University of Bonn, Br{\"u}hler Stra{\ss}e~7, 53175~Bonn, Germany}
\date[]{Accepted for publication: 20. Nov 2014, Physical Review E}
\begin{abstract}
We propose a straightforward extension of symbolic transfer entropy to enable the investigation of delayed directional relationships between coupled dynamical systems from time series.
Analyzing time series from chaotic model systems, we demonstrate the applicability and limitations of our approach.
Our findings obtained from applying our method to infer delayed directed interactions in the human epileptic brain underline the importance of our approach for improving the construction of functional network structures from data.
\end{abstract}
\maketitle
\begin{textblock*}{20cm}(3cm,27cm)
    Published as Phys.~Rev.~E \textbf{90}, 062706 (2014). Copyright 2014 by the American Physical Society.
\end{textblock*}
\section{Introduction}
Characterizing couplings between interacting systems plays an important role in numerous scientific fields, ranging from physics to the neurosciences~\cite{Pikovsky2001,Buzsaki2006,Osipov2007,Arenas2008,FellAxmacher2011,Sugihara2012,Schulz2013,Engel2013,TurkBrowne2013}.
Over the last years, a large number of linear and nonlinear analysis techniques has been proposed to reveal couplings from passive observations of the systems behavior, e.g., from time series of certain observables, and thus allow a data-driven quantification of the strength and direction of an interaction~\cite{Brillinger1981,Pikovsky2001,Boccaletti2002,Pereda2005,Hlavackova2007,Marwan2007,Lehnertz2009b,Lehnertz2011,Stankovski2012}.
Knowing interaction properties is important for the construction of functional network structures in diverse scientific fields~\cite{Boccaletti2006a, Arenas2008, Bullmore2009, Barabasi2011, Barthelemy2011, Sporns2011a, Bashan2012, Newman2012, Stam2012b, Lehnertz2014}.
Among these techniques, the information-theoretic concept of transfer entropy \cite{Schreiber2000} provides a model-free approach to characterizing directed interactions, because it can be viewed as transfers of information.
Transfer entropy is related to the concept of Granger causality \cite{Granger1969,Barnett2009} and to conditional mutual information \cite{Hlavackova2007}, and has widely been used
to distinguish the driving and responding elements and to detect asymmetry
in the interaction of subsystems in various scientific fields.
Since its invention, techniques that allow a data-driven estimation of transfer entropy are being steadily improved~\cite{Kaiser2002,Verdes2005,Hlavackova2007,Staniek2008,Kulp2009,Vakorin2009,Vlachos2010,Faes2011,Martini2011,Papana2011,Barnett2012,Stramaglia2012,Banerji2013,Kugiumtzis2013,Kugiumtzis2013b,Smirnov2013,Zuo2013}.
Among these improvements are methods that allow one to characterize information transfers at various time scales by incorporating delays~\cite{Nichols2005,Nichols2006,Overbey2009,Ito2011,Runge2012,Runge2012b,Naghoosi2013,Shu2013,Wibral2013}.
Knowing coupling delays is of importance as it allows for improved physical interpretations~\cite{Buenner2000a, Buenner2000b, Cimponeriu2004}.

In Ref.~\cite{Staniek2008}, symbolic transfer entropy has been proposed as a permutation analogue of transfer entropy and constitutes an efficient and conceptually simple way of robustly quantifying the dominating direction of flow of information between time series from observed data.
Using this approach, transfer entropy is estimated from the probabilities of ordinal patterns that are derived from the amplitude values of the time series via symbolization~\cite{Bandt2002}.
Symbolic transfer entropy has been used to study interactions in various disciplines ranging from quantum~\cite{Kowalski2010} and laser 
physics~\cite{Nian-Qiang2012} via neurology~\cite{Blain-Moraes2013}, cardiology~\cite{Jun2012} and anesthesiology~\cite{Ku2011,Jordan2013,Lee2013,Untergehrer2014} to the neurosciences~\cite{Martini2011,Zubler2015}.

Recently, an ordinal time series analysis technique has been introduced that detects the direction and the coupling delays of information exchange in coupled systems~\cite{Pompe2011}.
Here we follow this line of approach and propose a straightforward extension of symbolic transfer entropy, which we refer to as delayed symbolic transfer entropy.

This paper is organized as follows. In Sec.~\ref{sec_methods} we briefly recall the definition of symbolic transfer entropy before we present our extension to detect coupling delays and to quantify the dominating direction of flow of information.
In Sec.~\ref{sec_logistic_map} we present our numerical simulation studies that aim at demonstrating the applicability of our method and at exploring its limitations.
In Sec.~\ref{sec_eeg} we present our findings obtained from inferring delayed directed interactions in the human epileptic brain before we draw our conclusions in Sec.~\ref{sec_conclusion}.

\section{Symbolic transfer entropy and coupling delays}
\label{sec_methods}
\begin{figure*}
    \includegraphics[scale=0.85]{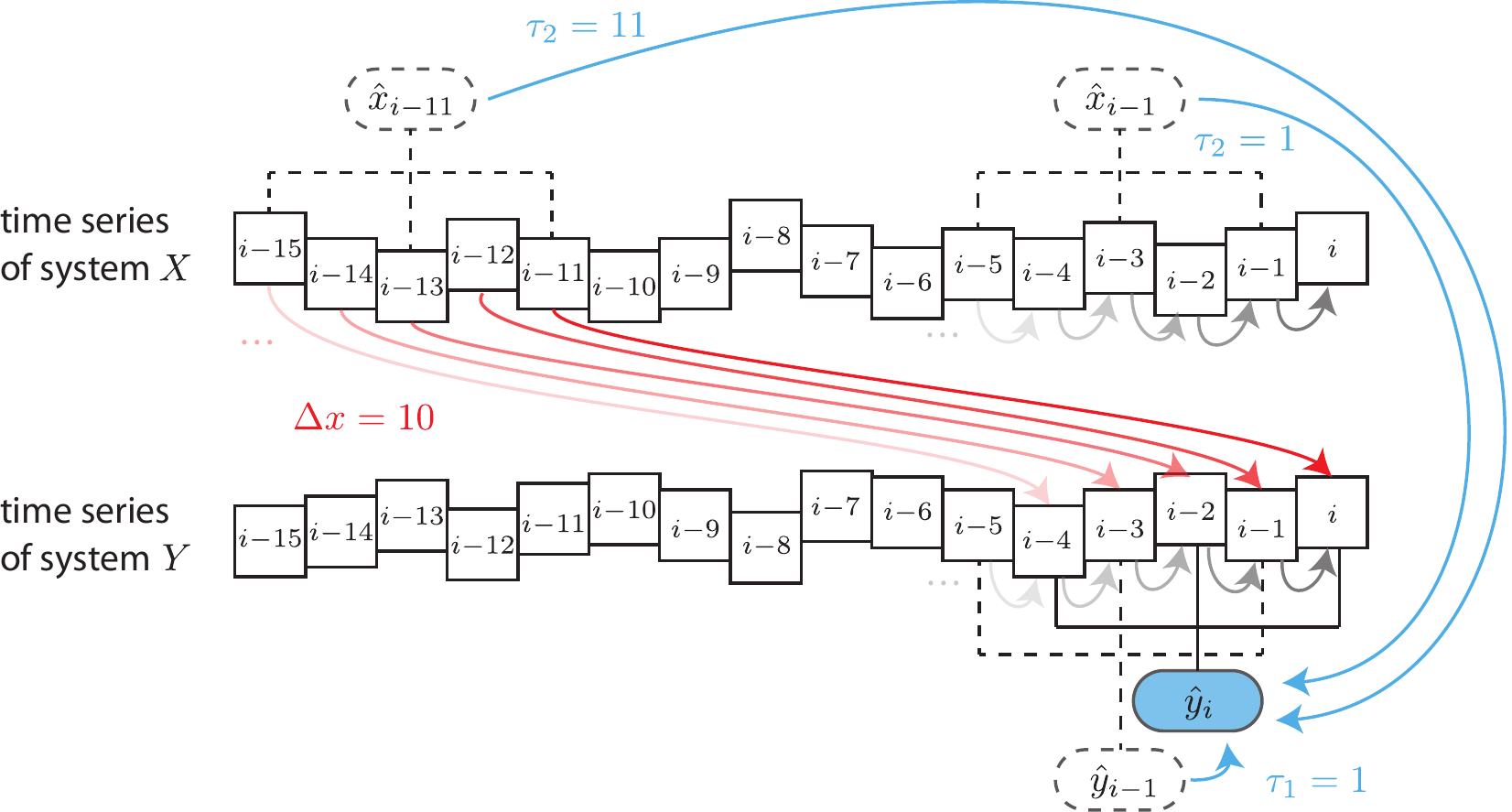}
    \caption{(Color online) Schematics of flow of information between two unidirectionally delay-coupled systems (\sysX drives \sysY) together with the procedure of symbolizing the time series and of estimating the delayed flow of information.
        The value of each element~$\stateX{i}$ ($\stateY{i}$) in the time series is encoded in the vertical position of the respective box.
        Flow of information from the system's own past to the current state~$i$ is indicated by gray arrows (transition from dark to light coloring indicate the loss of previous information, as the system evolves).
        The delayed flow of information (here $\delaycplX=10$) from states of system~\sysX to the current state of system \sysY is indicated by red arrows (transition from dark to light coloring as before).
        Blue arrows exemplify the procedure to estimate the flow of information:
        Arrows point from symbols, composed of exemplary previous states of either system~\sysX or~\sysY, to the actual symbol~$\symbY_i$, which is marked in blue.
        Here, the previous states $\symbX_{i-1}$ or $\symbX_{i-11}$  are \delayF time steps and $\symbY_{i-1}$ is \delayS time steps past the actual symbol~$\symbY_i$.
        The derivation of these permutation symbols is exemplarily shown for an embedding dimension~$\symbDim=3$ and lag~$\symbDelay=2$, e.g., $\symbX_{i-11} = (2, 1, 3)$.
    }
    \label{fig_coupling_scheme}
\end{figure*}

Let \stateX{i} and \stateY{i} with $i=1, \ldots, \numdp$ denote time series of observables of systems \sysX and \sysY.
Relating previous samples \stateX{i-1} and \stateY{i-1} in order to predict \stateY{i} allows for a quantification of the deviation from the generalized Markov property,
$\condprob{\stateY{i}}{\stateY{i-1}, \stateX{i-1}} \stackrel{!}{=} \condprob{\stateY{i}}{\stateY{i-1}}$, where \condprob{\cdot}{\cdot} denotes the conditional transition probability density.
If system~\sysX has no influence on system~\sysY, there is no deviation from the Markov property.
Transfer entropy quantifies the incorrectness of this assumption and is formulated as Kullback--Leibler entropy between \condprob{\stateY{i}}{\stateY{i-1}, \stateX{i-1}} and \condprob{\stateY{i}}{\stateY{i-1}}.
Transfer entropy is non-symmetric under the exchange of~\sysX and~\sysY.

In order to estimate the transition probabilities, the authors of Ref.~\cite{Staniek2008} proposed to use a symbolization technique with symbols that are derived from reordering the amplitude values of time series~\cite{Bandt2002}. Let \symbDelay~and~\symbDim denote the lag and embedding dimension, which have to be chosen appropriately for symbolization~\cite{Bandt2002,Staniek2007}, e.g., by making use of embedding theorems~\cite{Takens1981, Sauer1991, Kantz2003}.
Then \symbDim amplitude values
$
    s_i = \left( \stateX{i}, \stateX{i+\symbDelay}, \ldots, \stateX{i + \symbDelay(\symbDim - 1)} \right)
$
for a given, but arbitrary time $i$ are arranged in ascending order
$
\stateX{i + \symbDelay(\symbRank{1} - 1)} \leq \stateX{i + \symbDelay(\symbRank{2} - 1)} \leq \ldots \leq \stateX{i + \symbDelay(\symbRank{m} - 1)}
$
with rank~\symbRank{j} and $j \in \{1, \ldots, \symbDim\}$.
Equal amplitude values are arranged by their time index, i.e., such that 
$\symbRank{1} < \symbRank{2}$ if $\stateX{i+\symbDelay(\symbRank{1}-1)} = \stateX{i+\symbDelay(\symbRank{2}-1)}$.
This ensures that every $s_i$ is uniquely mapped onto one of the $\symbDim!$ possible permutations, and a permutation symbol is defined as
\begin{equation}
    \symbX_i := \left(\symbRank{1}, \symbRank{2}, \ldots, \symbRank{m} \right).
    \label{eq_symbolization}
\end{equation}
Relative frequencies of symbols provide an estimator for joint and conditional probabilities of the sequences of permutation indices.
With given symbol sequences $\symbX_i$ and $\symbY_i$, symbolic transfer entropy is defined as~\cite{Staniek2008}:
\begin{equation}
 \STE_\sysXY = \sum \joinprob{\symbY_i}{\symbY_{i-1}, \symbX_{i-1}} \log\frac{\condprob{\symbY_i}{\symbY_{i-1}, \symbX_{i-1}}}
          {\condprob{\symbY_i}{\symbY_{i-1}}},
  \label{eq_STE}
\end{equation}
where the sum runs over all symbols.
$\STE_\sysYX$ is defined in complete analogy.
$\STE_\sysXY$ is positive and explicitly non-symmetric under exchange of \sysX and \sysY since it measures the flow of information from \sysX to \sysY and not vice versa.
The difference $\STE_\sysYX - \STE_\sysXY$ provides an estimate for the dominating flow of information and thus for the dominating direction of interaction.

When analyzing empirical data, one often needs to account for delayed interactions (cf.~\reffig{fig_coupling_scheme}), where the flow of information from system \sysX to system \sysY needs some finite time $\delaycplX$ (and/or \delaycplY from \sysY to \sysX)~\cite{Mackey1977,deCarvalho2002,Mueller2003,Ermentrout2009,Batzel2011,Martin2014}.
Addressing this issue, we here extend \refeq{eq_STE} and allow for symbols in transition probabilities that are \delayS (\delayF) time steps past the actual symbol:
\begin{align}
    \begin{split}
        \DSTE_\sysXY (\delaySF) \! &:= \!\!\sum \joinprob{\symbY_i}{\symbY_{i-\delayS}, \symbX_{i-\delayF}} \!\log\frac{\condprob{\symbY_i}{\symbY_{i-\delayS}, \symbX_{i-\delayF}}} {\condprob{\symbY_i}{\symbY_{i-\delayS}}}\\
        \DSTE_\sysYX (\delaySF) \! &:= \!\!\sum \joinprob{\symbX_i}{\symbX_{i-\delayS}, \symbY_{i-\delayF}} \!\log\frac{\condprob{\symbX_i}{\symbX_{i-\delayS}, \symbY_{i-\delayF}}} {\condprob{\symbX_i}{\symbX_{i-\delayS}}}.
  \end{split}
  \label{eq_DSTE_single}
\end{align}
\delayS denotes the number of time steps into the systems' own past and \delayF the number of time steps into the past of the influencing system, i.e., the system from which we expect the flow of information.
We therefore did not interchange the parameters \delayS and \delayF in the definition of $\DSTE_\sysYX (\delaySF)$.
If there is a delayed flow of information from \sysX to \sysY (from \sysY to \sysX) and if $\delayF = \delaycplX$ ($\delayF = \delaycplY$), we expect \emph{delayed symbolic transfer entropy} $\DSTE_\sysXY (\delaySF)$ ($\DSTE_\sysYX (\delaySF)$) to attain highest values for all \delayS.
The use of the parameter \delayS may seem somewhat arbitrary, but we will see in the next section that $\DSTE_\sysYX (\delaySF)$ ($\DSTE_\sysXY (\delaySF)$) carries additional information for specific pairs $(\delaySF)$, which can assist in detecting delayed directed interactions in empirical data.
In the aforementioned definitions of entropies, we use a logarithm to base 2, thus entropies are given in bit.
\section{Applications}
\label{sec_applications}
\subsection{Delay-coupled logistic maps}
\label{sec_logistic_map}
In the following, we investigate the conditions under which delayed symbolic transfer entropy allows one to infer the coupling delays \delaycplX and \delaycplY and the direction of interaction.
Mimicking a typical experimental situation with a priori unknown coupling delays, we perform a parameter scan with ($\delaySF) \in \{1, \ldots, \delay_\text{max}\}$ in a range where we expect our maximum coupling delays.

We consider two delay-coupled logistic maps~\cite{Pompe2011} $f(\obsX) = r_x\,\obsX (1 - \obsX)$ with
\begin{equation}
    \begin{split}
        \stateX{i} &= f(g_\sysYX\!\!\mod 1),\\
        g_\sysYX  &= \couplY \stateY{i - 1 - \delaycplY} + (1 - \couplY) \stateX{i - 1},\\
        \stateY{i} &= f(g_\sysXY\!\!\mod 1),\\
        g_\sysXY &= \couplX \stateX{i - 1 - \delaycplX} + (1 - \couplX) \stateY{i - 1},
\end{split}
\end{equation}
where \couplX denotes the strength of coupling between systems \sysX and \sysY, and \couplY the respective strength between \sysY and \sysX.
For a slight mismatch of control parameter ($r_x = 3.9999, r_y = 3.9998$) as well as for given coupling strengths (\couplX,~\couplY) and coupling delays (\delaycplX,~\delaycplY), we generate 20 realizations of the system by randomly choosing the initial conditions (\stateX{0},~\stateY{0}) from the unit interval.
These time series consist of \numdp data points each after $10^4$ transients.
If not stated otherwise, we will report mean values of the delayed symbolic transfer entropies obtained from the 20~realizations of the coupled systems.
\subsubsection{General observations}
\begin{figure}[htb]
    \includegraphics{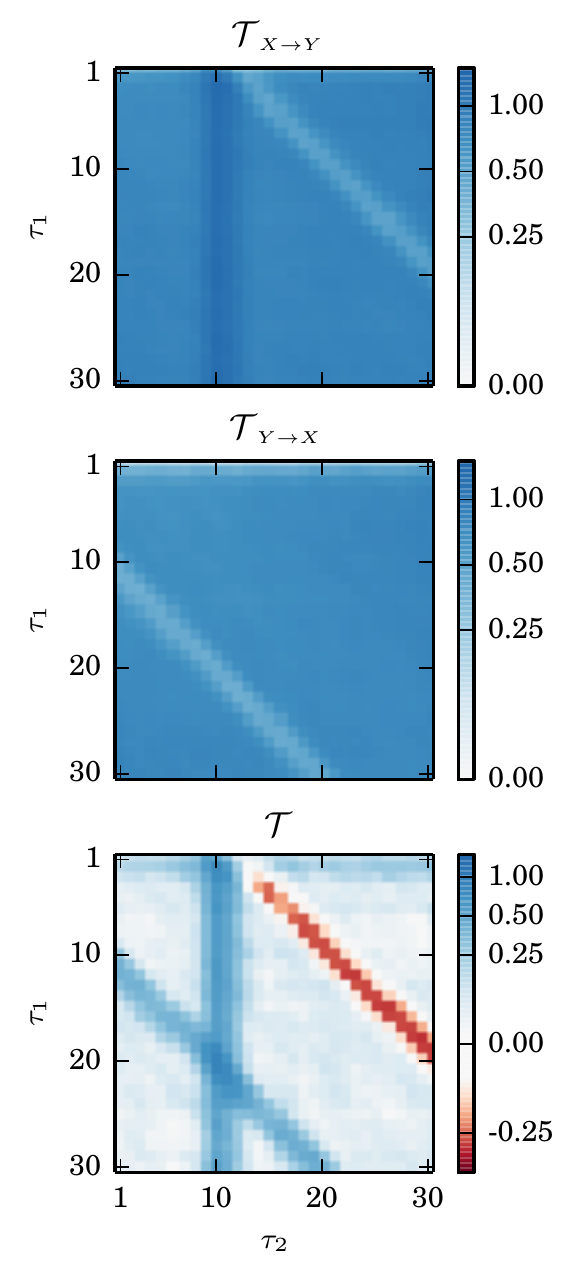}
    \caption{(Color online) Color-coded estimates of delayed symbolic transfer entropies $\DSTE_\sysXY (\delaySF)$ (top) and $\DSTE_\sysYX (\delaySF)$ (middle) and of the directionality index \DSTEd (bottom) for unidirectionally delay-coupled logistic maps
        ($\couplY=0, \couplX=0.45$, $\delaycplX=10$; embedding parameters $\symbDim = 3$, $\symbDelay=1$; $\numdp=100$ data points).
        Amplitude values are scaled linearly in $[-0.25, 0.25]$ and logarithmically otherwise.
    }
    \label{fig_logistic_te_single}
\end{figure}
In~\reffig{fig_logistic_te_single} we show, as an example, $\DSTE_\sysXY (\delaySF)$, $\DSTE_\sysYX (\delaySF)$, and the \emph{directionality index}
\begin{equation}
    \DSTEd:= \DSTE_\sysXY (\delaySF) - \DSTE_\sysYX (\delaySF)
    \label{eq_DSTE}
\end{equation}
for unidirectionally delay-coupled maps ($\couplY = 0, \couplX = 0.45$ and $\delaycplX=10$) obtained with embedding parameters $\symbDim = 3$ and $\symbDelay = 1$ and $\numdp=100$ data points.
When comparing findings for $\DSTE_\sysXY (\delaySF)$ with those for $\DSTE_\sysYX (\delaySF)$ there are two prominent effects:

First, if $\delayF = \delaycplX$ and for all \delayS, $\DSTE_\sysXY (\delaySF)$ attains highest values (up to three orders of magnitude larger than for other pairs (\delaySF), upper part of~\reffig{fig_logistic_te_single}), as expected and given our definition of delayed symbolic transfer entropy.
In the following, we will refer to this structure as \emph{resonance-like pattern}.

Second, we observe
$\DSTE_\sysXY (\delaySF)$ to attain lowest values, if $\delayF = \delayS + \delaycplX$ and $\delayS > 1$.
The same holds for $\DSTE_\sysYX (\delaySF)$, if $\delayF=\delayS - \delaycplX$ and $\delayS > \delaycplX$.
Interestingly, this strongly diminished (or even absent) flow of information for these \emph{secondary diagonals} in the plots shown in~\reffig{fig_logistic_te_single} (upper part: upper diagonal; middle part lower diagonal) also provide information about delay and direction of interaction.
For exactly these pairs (\delaySF), the delayed flow of information has just taken place and thus these past states of system~\sysX and~\sysY provide almost the same amount of information for the current state of system~\sysY (cf.~\reffig{fig_coupling_scheme}).
This leads to only a few combinations of symbols contributing to the transition probabilities.
Consequently, the ratio of conditional probabilities in~\refeq{eq_DSTE_single} approaches 1 and thus $\DSTE_\sysXY (\delaySF)$ approaches~0 (the same holds for  $\DSTE_\sysYX (\delaySF)$).
For all other pairs (\delaySF) not considered yet (referred to as \emph{background} in the following), the delayed flow of information $\DSTE_\sysXY (\delaySF)$ ($\DSTE_\sysYX (\delaySF)$) only approaches~0 for an increasing number of data points and for appropriately chosen embedding parameters (see above).
For a wide range of coupling strengths differentiability of the secondary diagonals from the background (i.e., the difference to the background) is thus best for small numbers of data points accompanied by non-optimally chosen embedding parameters as is often the case when analyzing empirical data.
As an example, for embedding parameters $\symbDim=3$ and $\symbDelay=1$, which are optimal for the system considered here, differentiability is almost~0 for $\numdp=10^5$ but increases almost exponentially with decreasing the number of data points to~$\numdp=10^2$.

The directionality index \DSTEd, as defined here, provides information about delay and direction of interaction.
If we exchange system~\sysY for \sysX, this leads to a change of sign of values of the directionality index \DSTEd, since the resonance-like pattern and the upper secondary diagonal can now be observed with $\DSTE_\sysYX (\delaySF)$ and the lower secondary diagonal with $\DSTE_\sysXY (\delaySF)$.

Note that for $\delayS = \delayF = 1$, delayed symbolic transfer entropies correspond to the non-delayed ones and fail to correctly detect the delayed coupling, as expected (cf.~\reffig{fig_coupling_scheme}).
Since $\DSTEd \approx 0$, this also applies for the direction of interaction, independent on coupling strength, number of data points, and embedding parameters (at least for the cases considered in this section).
\subsubsection{Influence of the number of data points \numdp and the embedding parameters \symbDim and \symbDelay}
For unidirectionally coupled maps ($\couplY = 0$) with coupling delays $(\delaycplX, \delaycplY) \in \{1, \ldots, 25\}$ and coupling strengths $\couplX \in [0, 0.7]$, we generate time series consisting of $\numdp \in \{10^2, \ldots, 10^6\}$ data points and estimate \DSTEd for embedding dimensions $\symbDim \in \{2, \ldots, 5\}$ and lags $\symbDelay \in \{1, \ldots, 5\}$ (cf.~\cite{Takens1981,Sauer1991,Schuermann1996,Runge2012}).
\begin{figure*}
    \includegraphics{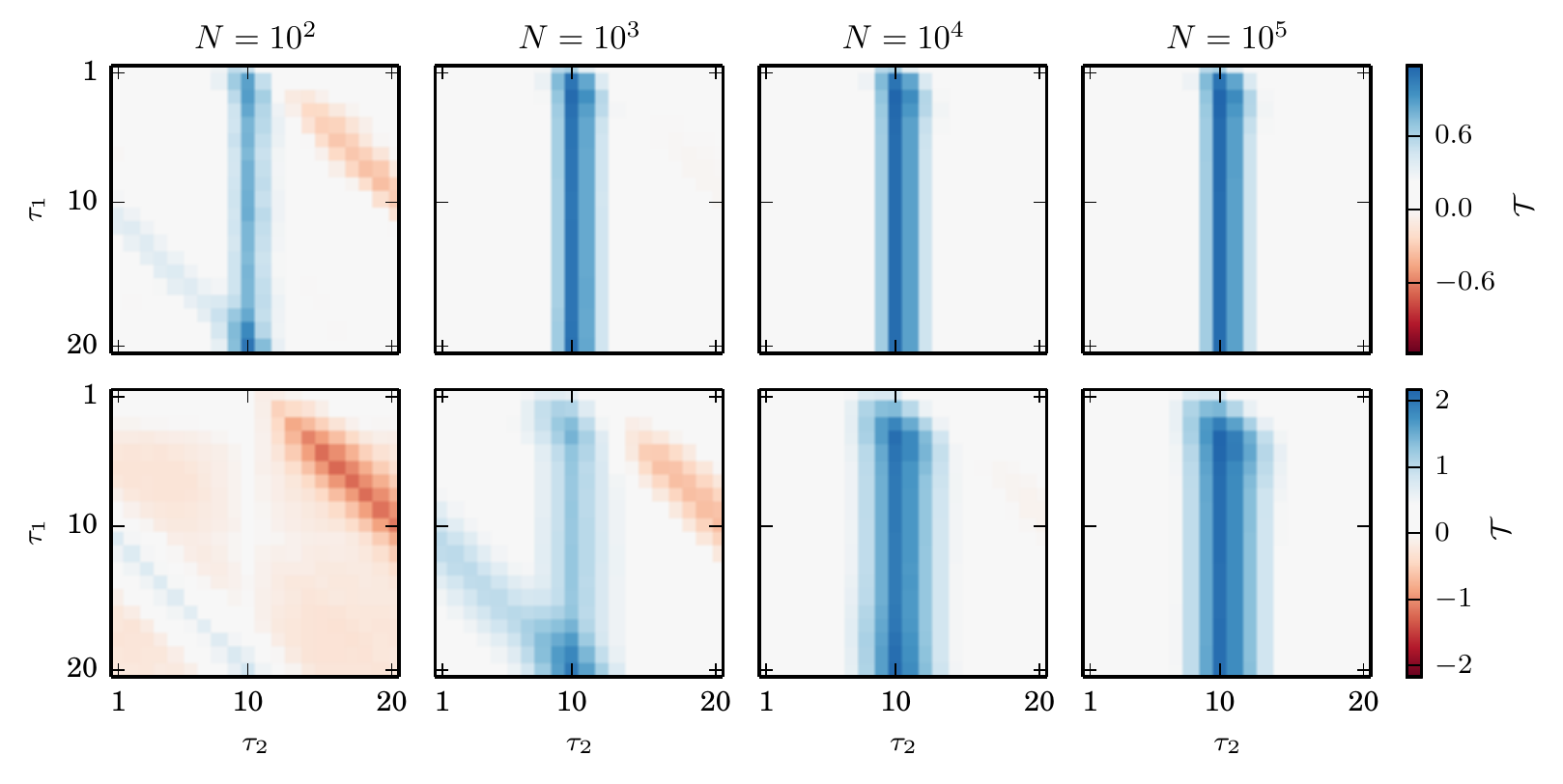}
    \caption{(Color online) Color-coded estimates of the directionality index \DSTEd for unidirectionally delay-coupled logistic maps with delay $\delaycplX=10$ and coupling strength $\couplX=0.45$.
        Embedding parameters: (top) $\symbDim=3, \symbDelay=1$, (bottom) $\symbDim=4, \symbDelay=1$.
        Left to right: increasing number of data points~\numdp.
        Positive (negative) values of \DSTEd indicate the driving (responding) behavior of system~\sysX.}
    \label{fig_logistic_numdp}
\end{figure*}
In \reffig{fig_logistic_numdp} we demonstrate exemplarily, how inference depends on the number of data points~\numdp and on the embedding dimension~\symbDim.
When decreasing~\numdp, the amplitude of the resonance-like pattern decreases and, dependent on the chosen embedding dimension~\symbDim, even vanishes.
Instead, for smaller~\numdp, the secondary diagonals can be observed.
As a rule of thumb (and at least for the systems investigated here), $\numdp \approx 10 ^ {\symbDim - 1}$ marks the border above which delay and direction of interactions can be inferred from the resonance-like pattern.
Below this border but above a lower bound which depends on system properties, the same information can be inferred from the secondary diagonals.
The width of the patterns increases linearly with the embedding dimension~\symbDim.
This broadening can be attributed to the applied symbolization technique~\cite{Vlachos2010,Pompe2011,Kugiumtzis2012}, since the overlap of symbols grows linearly with the embedding dimension~\symbDim (cf.~\reffig{fig_coupling_scheme}).
\begin{figure*}
    \includegraphics{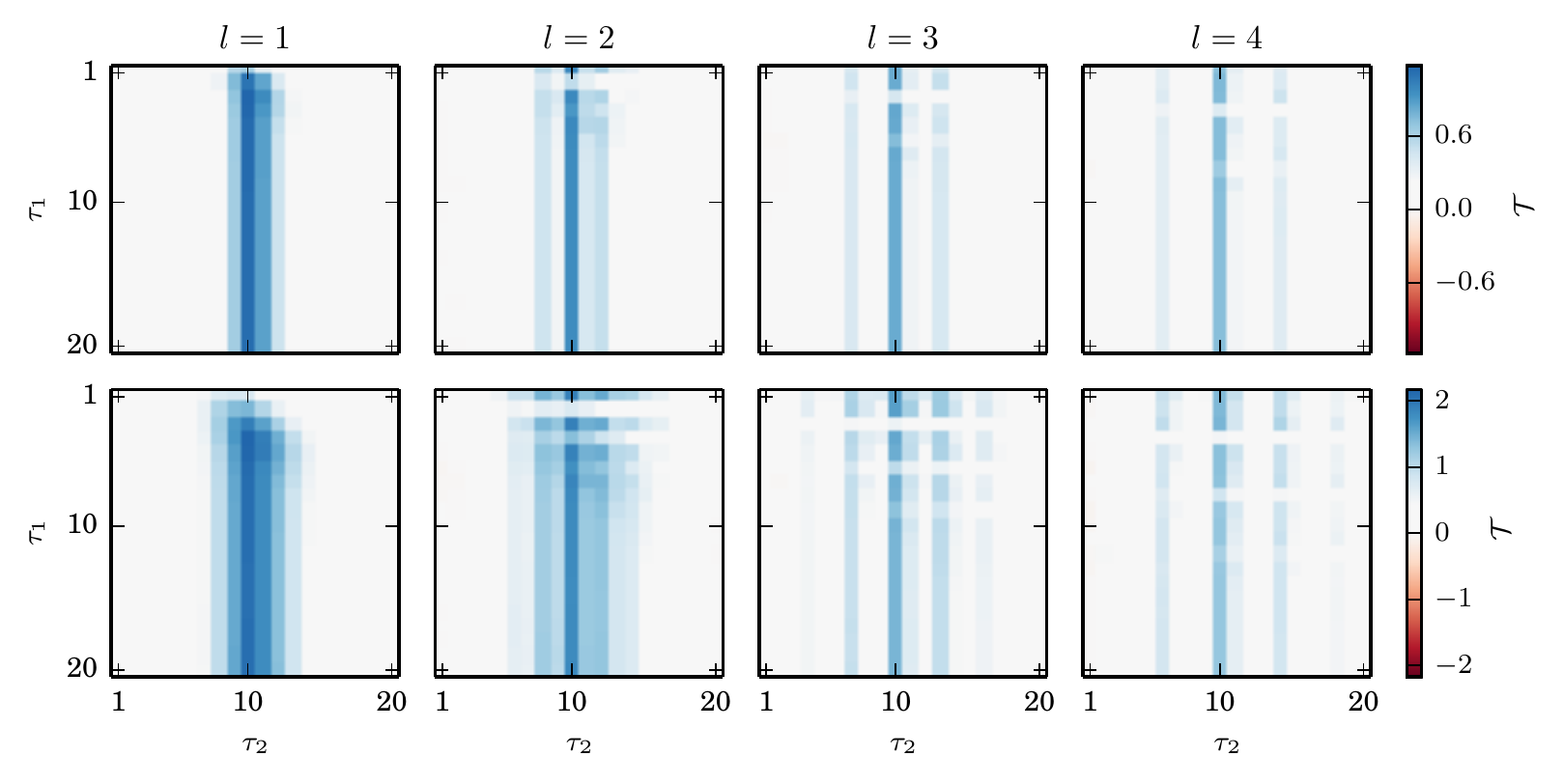}
    \caption{(Color online) Color-coded estimates of the directionality index \DSTEd for unidirectionally delay-coupled logistic maps with delay $\delaycplX=10$ and coupling strength $\couplX=0.45$ estimated with $\numdp=10^{\symbDim + 1}$ data points.
        Embedding dimension: (top) $\symbDim=3$ (bottom) $\symbDim=4$.
        Left to right: increasing embedding lag~\symbDelay.
        Positive (negative) values of \DSTEd indicate the driving (responding) behavior of system \sysX.}
    \label{fig_logistic_embedding}
\end{figure*}

The influence of the embedding lag~\symbDelay is demonstrated exemplarily in~\reffig{fig_logistic_embedding} for the resonance-like pattern.
For given \numdp~and~\symbDim and with $\symbDelay > 1$, highest values of \DSTEd can still be observed for $\delayF=\delaycplX$ and for all~\delayS, but we observe additional resonance-like patterns, if $\delayF \approx \delaycplX \pm j \symbDelay$, for $j \in \{0, \ldots, \symbDim - 1\}$, however with lower values of \DSTEd\@.
Within these patterns, \DSTEd attains lower values if $\delayS \in \{\symbDelay, \ldots, (\symbDim-1) \symbDelay\}$, which again is linked to the applied symbolization technique.
For these conditions, permutation symbols share up to $\symbDim -1$ amplitude values and are therefore not independent.
Analogous observations hold for the secondary diagonals, and we obtained similar findings for other coupling delays.
\subsubsection{Impact of strength and type of coupling}
\label{sec_couplingstrength}
\begin{figure*}
    \includegraphics[scale=0.97]{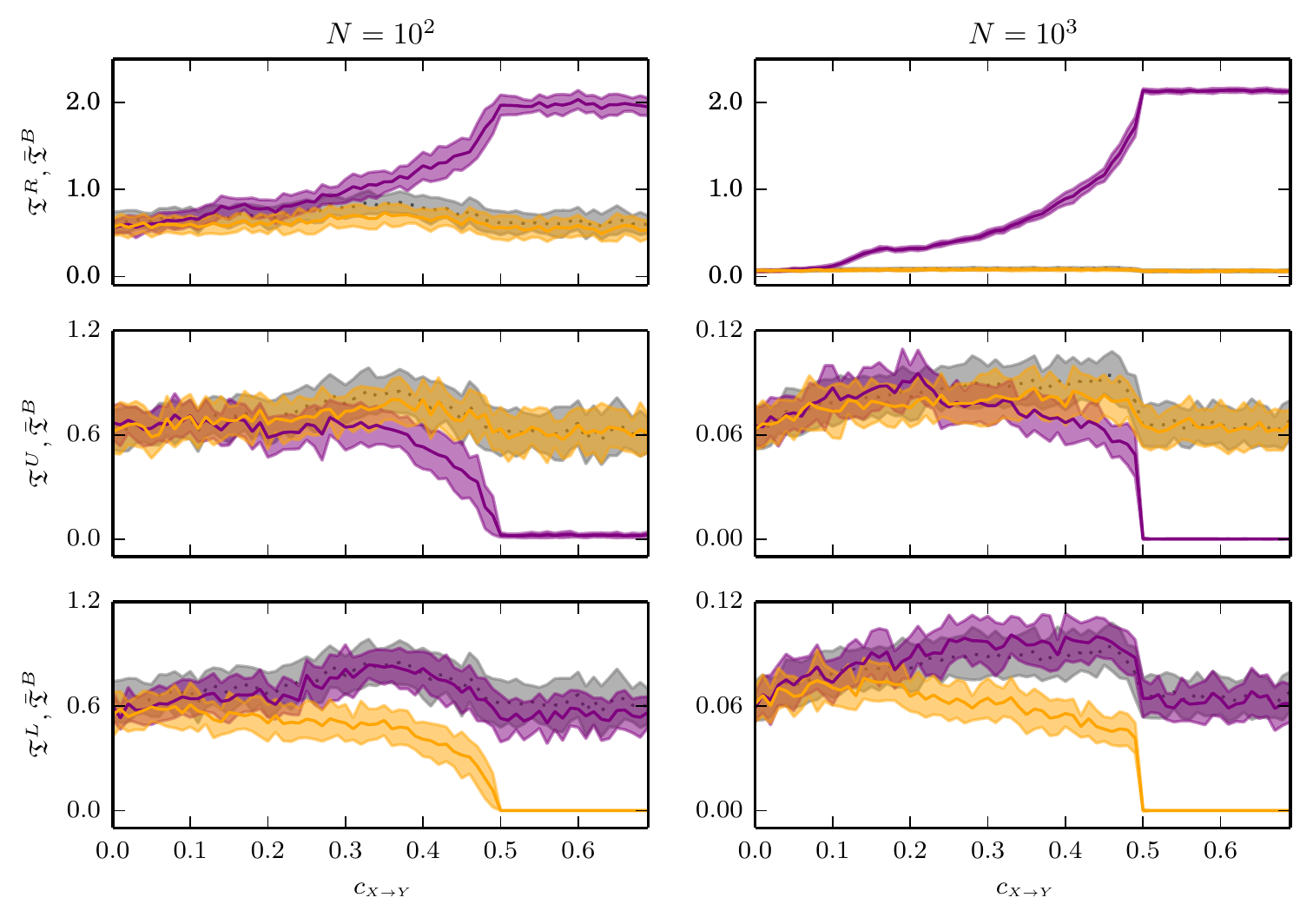}
    \caption{(Color online) Means and standard deviations of delayed symbolic transfer entropies for the directions $\sysX \to \sysY$ (purple line and shaded area) and $\sysY \to \sysX$ (orange line and shaded area) depending on the coupling strength~\couplX for 20 realizations of delay-coupled logistic maps with delay $\delaycplX=10$ and $\numdp = 100$ (left) and $\numdp=1000$ data points (right).
        Embedding parameters: $\symbDim=3$~and~$\symbDelay=1$.
        Upper row: estimates $\DSTEs^R$ from the resonance-like pattern with $\delayS=11$ and $\delayF = \delaycplX = 10$;
        middle row: estimates $\DSTEs^U$ from the upper secondary diagonal (cf.~\reffig{fig_logistic_te_single}) with $\delayS=11$ and $\delayF = 21$;
        lower row: estimates  $\DSTEs^L$ from the lower secondary diagonal with $\delayS=11$ and $\delayF = 1$.
        In all plots, averaged estimates from the background $\bar \DSTEs^B$ (cf.~\reffig{fig_logistic_te_single}) with $\delayS = \delayF = 25$ are shown in gray.}
    \label{fig_logistic_coupling_comparison}
\end{figure*}
In the following, we fix the embedding parameters ($\symbDim=3$ and $\symbDelay=1$)
and investigate the impact of type and strength of coupling on the inference of delayed directed interactions.
For unidirectional couplings with delay $\delaycplX=10$ we show, in~\reffig{fig_logistic_coupling_comparison}, the dependence of delayed symbolic transfer entropies on the coupling strength and for two numbers of data points.
We make use of a priori knowledge for which pairs (\delaySF) we can expect the resonance-like pattern ($\DSTEs^R_\sysXY = \DSTE_\sysXY (11, 10)$, upper row) and the secondary diagonals ($\DSTEs^U_\sysXY = \DSTE_\sysXY (11, 21)$, middle row, and $\DSTEs^L_\sysXY = \DSTE_\sysXY (11, 1)$, lower row); the assignments for the opposite direction $\sysY \to \sysX$ are analogous.
In addition, we show the mean of both directions for a pair (\delaySF), for which there is no pattern, i.e., from the background: $\bar \DSTEs^B = (\DSTE_\sysXY(25, 25) + \DSTE_\sysYX (25, 25)) / 2$.

For a larger number of data points (here $\numdp = 10^3$) the flow of information can be inferred even for small coupling strengths ($\couplX \approx 0.1$) and differentiability of $\DSTEs^R_\sysXY$ from the background increases with increasing coupling strength up to a maximum at $\couplX \approx 0.5$, for which the systems are lag-synchronized (\reffig{fig_logistic_coupling_comparison}, upper row).
For larger coupling strengths differentiability remains at its maximum value.
For the opposite direction, there is no flow of information and we obtain $\DSTEs^R_\sysYX \approx \bar\DSTEs^B$ for all coupling strengths, as expected.
For smaller number of data points (here $\numdp = 10^2$), standard deviations of estimates are generally enlarged, as expected.
In addition, mean values of estimates are increased, and the increase is stronger for $\bar \DSTEs^B$ (and $\DSTEs^R_\sysYX$) than for $\DSTEs^R_\sysXY$.
Inference of flow of information is thus diminished and restricted to coupling strengths $\couplX \gtrapprox 0.4$.

Making use of information gained from the upper secondary diagonal (\reffig{fig_logistic_coupling_comparison}, middle row), the deviation of $\DSTEs^U_\sysXY$ from $\bar \DSTEs^B$ for $\couplX \gtrapprox 0.45$ and~$\numdp = 10^3$ also indicates the inference of flow of information.
For~$\numdp = 10^2$, inference can already be achieved for $\couplX \approx 0.4$.
Again, $\DSTEs^U_\sysYX \approx \bar \DSTEs^B$ for all coupling strengths and number of data points.
Note, however, that both means and standard deviations of estimates are increased by one order of magnitude when decreasing~\numdp from $10^3$~to~$10^2$.
An even better inference of flow of information can be achieved from information gained from the lower secondary diagonal (\reffig{fig_logistic_coupling_comparison}, bottom row).
Although similar observations can here be made for means and standard deviations of estimators, $\DSTEs^L_\sysYX$ (and not $\DSTEs^L_\sysXY$, given our definitions; see~\refeq{eq_DSTE}) deviates clearly from $\bar \DSTEs^B$ for coupling strengths $\couplX \gtrapprox 0.25$ for both numbers of data points considered here.

Summarizing these findings, in the case of smaller number of data points, directed interactions can be inferred for a larger range of coupling strengths with information from the lower secondary diagonal ($\DSTEs^L_\sysYX$) than from the resonance-like pattern ($\DSTEs^R_\sysXY$).

For bidirectionally delay-coupled maps, similar observations can be made (data not shown here), as long as the coupling delays \delaycplX~and~\delaycplY as well as the coupling strengths \couplX~and~\couplY are not identical.
Even for the case $\delaycplX = \delaycplY$ the dominating delayed flow of information can be inferred, if the coupling strengths are sufficiently different (cf.~\reffig{fig_logistic_coupling_comparison}).
As before, inference is influenced by alterations of the patterns (the resonance-like pattern and the secondary diagonals) related to the choice of embedding parameters necessary for the applied symbolization technique.
\subsubsection{Influence of noise}
\label{sec_noise}
\begin{figure*}
    \includegraphics[scale=0.97]{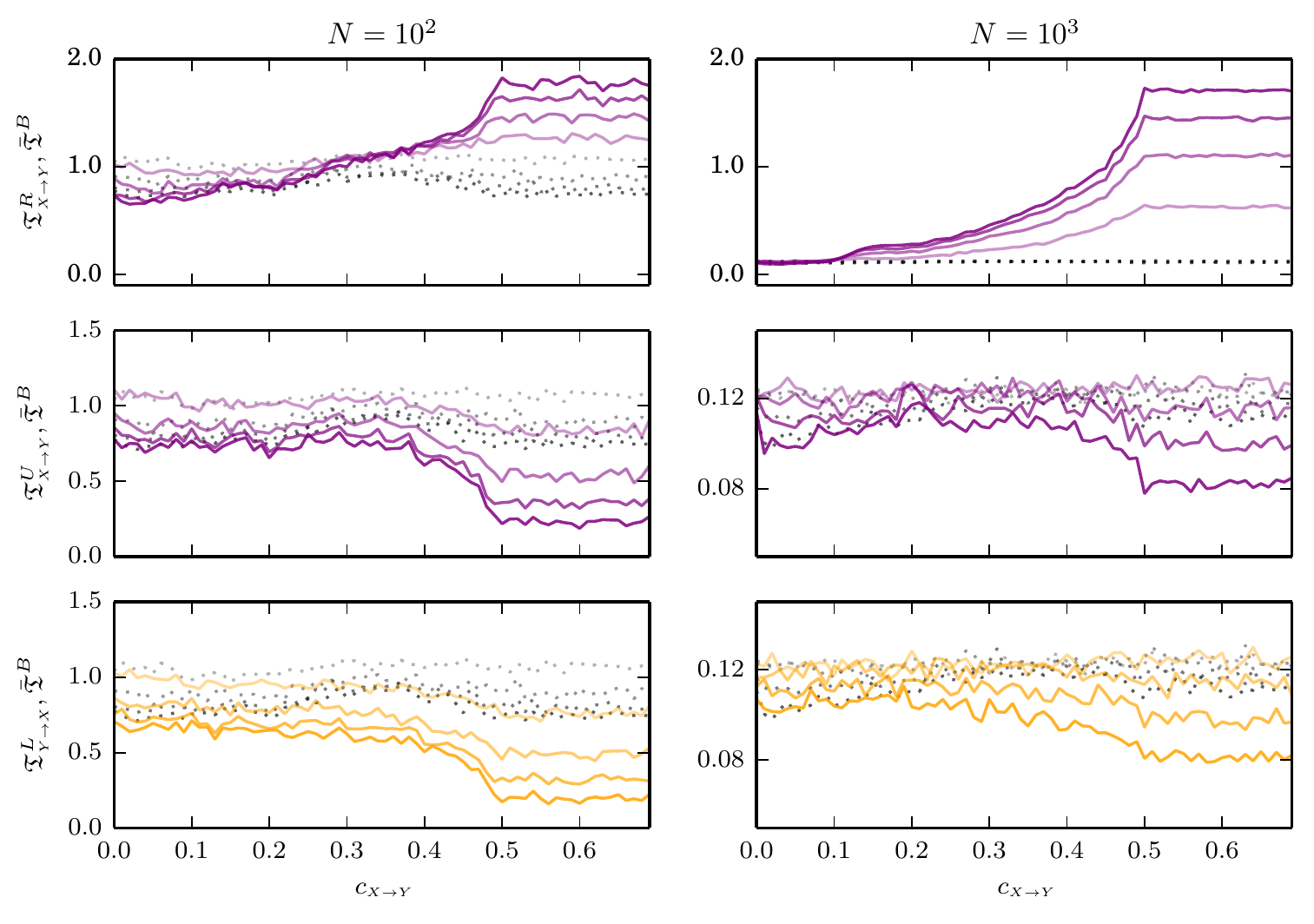}
    \caption{(Color online) Means of delayed symbolic transfer entropies for the directions $\sysX \to \sysY$ (purple) and $\sysY \to \sysX$ (orange) depending on the coupling strength~\couplX.
        Twenty realizations of noisy delay-coupled logistic maps with delay $\delaycplX=10$ and $\numdp = 100$ (left) and $\numdp=1000$ data points (right).
        The transitions from dark to light coloring encodes a decreasing signal-to-noise ratio $(128, 32, 8, 2)$.
        Embedding parameters and choice of pairs (\delaySF) for delayed symbolic transfer entropies~\DSTEs as in~\reffig{fig_logistic_coupling_comparison}.
    }
    \label{fig_logistic_noiselevel}
\end{figure*}
Next we estimate the performance of our method, particularly with respect to the analysis of empirical data, by investigating the influence of noise on the inference of delayed directional couplings.
For unidirectional couplings with delay $\delaycplX=10$, we add noise to the time series \stateX{i} of the driver and to the time series \stateY{i} of the responder, and estimate \DSTEd for signal-to-noise ratios $\SNR \in [1, 128]$ ($\SNR = \frac{\sigma_s}{\sigma_n}$, where $\sigma_s$ and $\sigma_n$ denote the standard deviations of the noise-free and the noise-contaminated time series).
We use different types of noise as well as different noise-contamination schemes.
With Gaussian $\delta$-correlated noise, we simulate measurement errors,
and with the concept of surrogates~\cite{Schreiber1996a}, we generate in-band noise from the original time series, thus mimicking observational noise.
The surrogate time series have a power spectrum and a distribution of amplitude values that are identical to those of the original time series.
With each of these types of noise we contaminate both time series, \stateX{i}~and~\stateY{i}, using either the same~\SNR \emph{(symmetric noise contamination)} or different~\SNR for the driver and responder \emph{(asymmetric noise contamination)}.
The latter contamination scheme is more likely in field applications
and is known to affect various time series analysis techniques aiming at an inference of the direction of interactions~\cite{Smirnov2003,QuianQuiroga2000,Chicharro2009, Albo2004, Nolte2004}.

In~\reffig{fig_logistic_noiselevel}, we show exemplary findings for a symmetric contamination with in-band noise.
For various \SNR we plot the dependence of delayed symbolic transfer entropies on the coupling strength and for different number of data points.
As in the previous subsection, we restrict ourselves to the pairs (\delaySF) for which we can expect the correct direction of flow of information from the resonance-like pattern ($\DSTEs^R_\sysXY$) and from the secondary diagonals ($\DSTEs^U_\sysXY$ and $\DSTEs^L_\sysYX$).

As expected, differentiability of all estimators of flow of information from the background $\bar \DSTEs^B$ decreases with an decreasing signal-to-noise ratio.
Likewise, the range of coupling strengths for which directed interactions can be inferred shrinks with decreasing the signal-to-noise ratio and is shifted towards higher coupling strengths.
For a smaller number of data points, the inference of flow of information and with this the direction of interaction gained from the secondary diagonals ($\DSTEs^U_\sysXY$ and $\DSTEs^L_\sysYX$) is more robust to noise contaminations than for a larger number of data points.
As expected, the opposite is true for the inference gained from the resonance-like pattern ($\DSTEs^R_\sysXY$).
We obtained similar findings for the other types of noise and contamination schemes.

\subsubsection{Summary}
Taking advantage of the conceptual simplicity, efficiency, and robustness of symbolic transfer entropy, we demonstrated that our extension allows to infer of delayed directed interactions.
Our method provides information about delay and direction of couplings even for smaller number of data points and, moreover, for the case of a non-optimal choice of embedding parameters used for the symbolization.
This renders delayed symbolic transfer entropy attractive for the analysis of empirical data.
\subsection{Inferring delayed directed interactions in the human epileptic brain}
\label{sec_eeg}

\begin{figure*}
    \begin{minipage}[c]{0.49\linewidth}
        \includegraphics[width=3.37in]{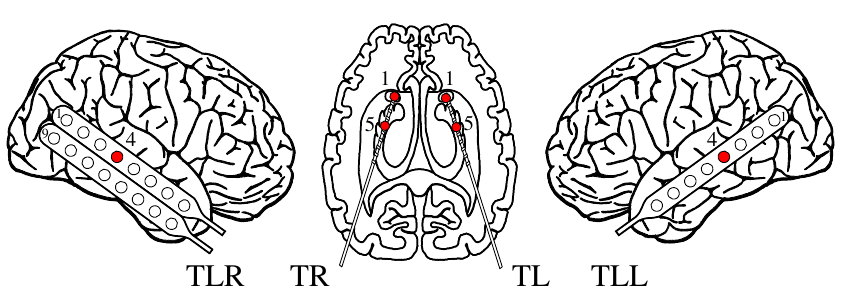}
    \end{minipage}
    \begin{minipage}[c]{0.49\linewidth}
        \includegraphics[width=3.37in]{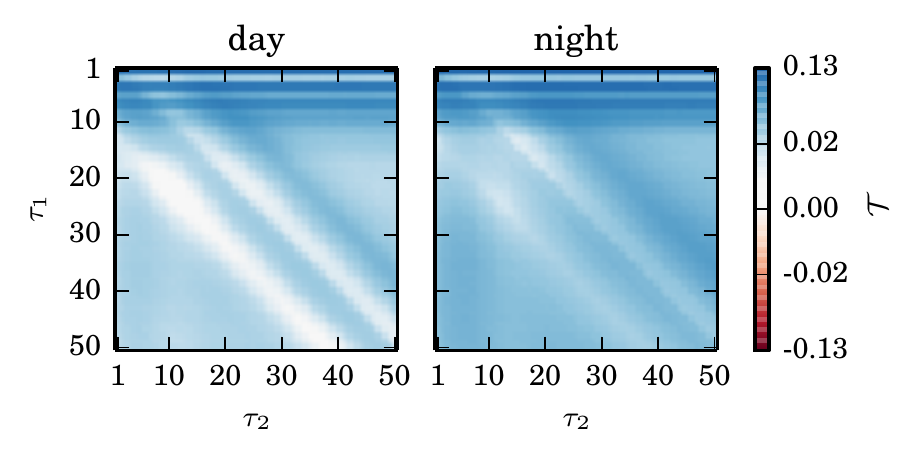}
    \end{minipage}
    \includegraphics[width=3.37in]{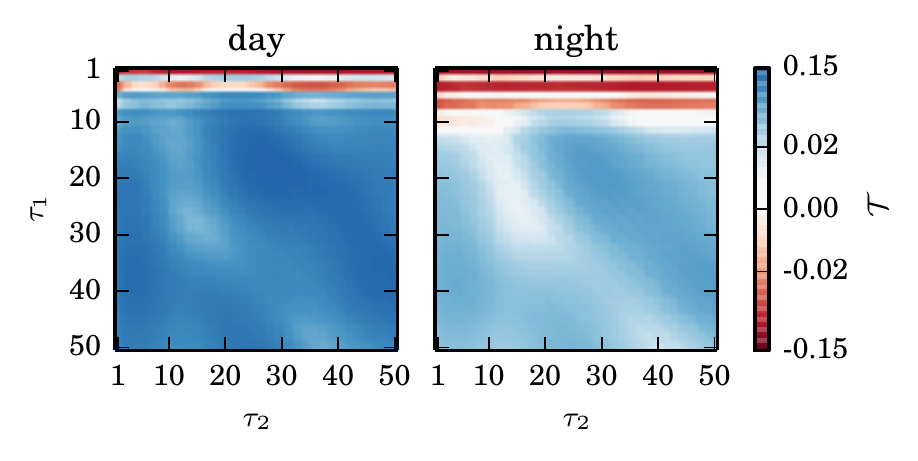}
    \includegraphics[width=3.37in]{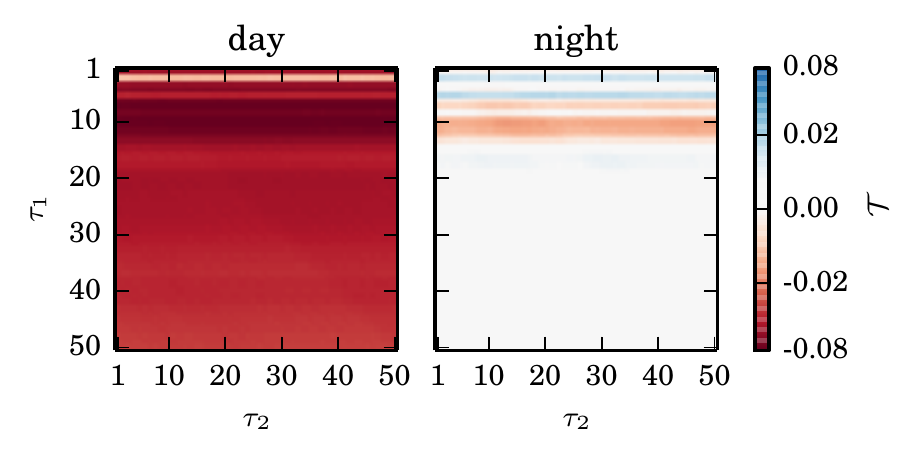}
    \caption{(Color online) Upper left: Schematics of electrode strips placed over the left and right temporal lateral neocortex and of bilateral intrahippocampal depth electrodes. Recording sites that were used for analyses are marked red.
        Upper right to lower right: Color-coded estimates of the mean directionality index \DSTEd for interactions between various brain regions estimated from intracranial EEG recorded during day times and night times. Interactions between anterior and posterior sites within the non-epileptic mesial temporal brain structures (TL01-TL05, upper right), between homologous sites within the epileptic brain hemisphere (TR01-TR05, lower left) and between regions in the left and right temporal lateral neocortex (TLL04-TLR04, lower right). The horizontal stripes that can be observed for \delayS taking on integer multiples of the embedding lag \symbDelay can be related to the applied symbolization (cf.~\reffig{fig_logistic_embedding}).}
    \label{fig_fielddata_results}
    \label{fig_fielddata_implscheme}
\end{figure*}

In this section, we apply our method to check whether consistent delayed directed interactions between brain regions can be inferred from long-lasting, multichannel electroencephalographic (EEG) recordings.
The EEG was recorded from an epilepsy patient using electrodes implanted under the skull, hence with high signal-to-noise ratio, prior to surgical treatment of a focal epilepsy.
The patient had signed informed consent that her/his clinical data might be used and published for research purposes.
The study protocol had previously been approved by the ethics committee of the University of Bonn.
We here consider EEG recordings from strip electrodes (8 or 16 contacts) placed onto the cortex and from a pair of needle-shaped depth electrodes with 10 contacts each, implanted into deeper structures of the brain (see upper left part of~\reffig{fig_fielddata_implscheme}).
Data were sampled at \unit[200]{Hz} (sampling interval $\Delta t = $ \unit[5]{ms}) using a 16~bit analog-to-digital converter and filtered within the frequency band~\unit[1--45]{Hz}.

For our analyses, we consider a continuous recording of \unit[36]{h} duration during the seizure-free interval, which covered different physiologic and pathophysiologic states of the patient.
Here we restrict ourselves to EEG data from six recording sites (see upper part of~\reffig{fig_fielddata_implscheme}): two from within the epileptic focus (TR01 and TR05), one remote site on the same brain hemisphere (TLR04), and three from homologous positions on the other brain hemisphere (TL01, TL05, and TLL04).
A widely used approach to analyze the dynamics of non-stationary systems is to perform the analysis in sliding windows with a duration, for which the dynamics can be regarded as approximately stationary.
For the EEG, the duration of such a window typically amounts to \unit[20]{s} duration~\cite{Blanco1995}.
Using this approach, we perform---for each combination of pairs of recording sites---a time resolved estimation of delayed symbolic transfer entropies from non-overlapping EEG segments of \unit[20.48]{s} duration (corresponding to 4096 data points).
Following Ref.~\cite{Staniek2008}, we set embedding parameters to $\symbDim=5$ and $\symbDelay=3$.

Since time delays in the human brain can vary considerably, depending on brain regions and functions, and may reach up to \unit[200]{ms} \cite{Nunez1995}, we estimate \DSTEd with $(\delaySF) \in \{\Delta t,\ldots, 50\Delta t\} = \{\unit[5]{ms}, \ldots, \unit[250]{ms}\}$.
Moreover, by time-averaging separately over all windows for data recorded during day and during night times, we check whether major delays as well as preferred directed interactions can be identified and whether delay and direction depend on the state of consciousness (awake vs.\@ asleep).
In~\reffig{fig_fielddata_results}, we show the mean directionality indices~\DSTEd separately for data from day and night times for three exemplary pairs of recording sites.
In general, we do not observe the resonance-like patterns, which is to be expected given the number of data points and embedding parameters.
For some cases, however, we observe secondary diagonals, from which we can extract information about delay and direction of an interaction.
In particular, we observe a consistent driving with an average delay of \unit[60]{ms} (\unit[55--65]{ms}) from posterior (position TL05) to anterior sites (position TL01) within the non-epileptic (left) mesial temporal brain structures both during day and night times (upper right part of figure).
For homologous recording sites within the epileptic (right) mesial temporal brain structures a similar directed driving with an average delay of \unit[50]{ms} (\unit[35--65]{ms}) can be observed for data recorded during night times (lower left part of figure).
This delay is comparable to findings gained from analyses of propagation of specific patterns during seizures~\cite{Gotman1987,Bertashius1991,Alarcon1996}.
Identifying a delay for data recorded during day times, however, is more demanding, possibly due to multiple delays (which may be associated with the epileptic process).
Interestingly, for $\delayS = \delayF = 1$, for which \DSTEd corresponds to the non-delayed directionality index, we observe the direction of driving to be reversed, i.e., from anterior to posterior sites.
Our findings for long-ranged interactions between regions in the left and right temporal lateral neocortex (lower right part of figure) also point to multiple delays, and it remains to be shown whether they differ from those obtained for the short-ranged interactions within the epileptic focus.
For data recorded during day times, the brain region in the right temporal lateral neocortex constantly drives the homologous brain region in the left hemisphere.
This unidirectional driving vanishes for data recorded during night times, and we can only speculate whether this is due to, e.g., a bidirectional interhemispheric driving or a diminished interhemispheric interaction during sleep (cf.~\cite{Duckrow2005,Bertini2009}).

\section{Conclusions}
\label{sec_conclusion}
We have proposed a straightforward extension of symbolic transfer entropy~\cite{Staniek2008} that enables the inference of delayed directional relationships between coupled dynamical systems from time series. 
With numerical examples, which are representative of interacting chaotic systems contaminated with noise, we have exemplified the applicability of our approach and have shown that delay and direction of an interaction can be inferred with delayed symbolic transfer entropy even for smaller number of data points and, moreover, with non-optimally chosen parameters for the applied symbolization technique~\cite{Bandt2002}.
Applying our method to infer delayed directed interactions in the human epileptic brain, we could show that major interaction delays can be identified, particularly from short-ranged interactions, and that these delays are influenced by the pathophysiology and by physiologic states of the brain.
Moreover, we could also show, that not taking into account possible delays in interactions can lead to a possibly erroneous inference of the direction of interactions.
Our approach can thus help to avoid misinterpretations and to further improve the construction of functional network structures from data~\cite{Lehnertz2014}.

At present, our approach requires estimating the directionality index \DSTEd with parameters ($\delaySF$) in a range where we expect maximum coupling delays.
Although a more direct detection of coupling delays would be preferable, we note that the identification of delayed directed interactions from time series (4096 data points, embedding dimension $\symbDim=5$) for all $(\delaySF) \in \{1, \ldots, 50\}$ can be performed in about \unit[60]{s} on a \unit[2.5]{GHz} CPU core due to the underlying conceptual simplicity, efficiency, and robustness of symbolic transfer entropy.

\section*{Acknowledgments}
We are grateful to Gerrit Ansmann, Christian Geier, Stephan Porz, and Alexander Rothkegel for critical comments on earlier versions of the manuscript.
This work was supported by the Deutsche Forschungsgemeinschaft (Grant No: LE 660/5-2).

\end{document}